\documentclass[aps,pre,showpacs,noshowkeys,amsmath,amssymb,amsfonts,superscriptaddress,longbibliography,reprint]{revtex4-1}
\usepackage[english]{babel}

\usepackage{graphicx}
\usepackage{bm}
\usepackage{physics}
\usepackage{mathtools}
\usepackage{gensymb}
\usepackage{soul}
\usepackage{comment}

\setcitestyle{super}
\usepackage{caption}
\usepackage{subcaption}
\DeclareCaptionLabelSeparator{bar}{~\rule[-0.4ex]{0.2ex}{1em}~}
\DeclareCaptionLabelFormat{subfor}{\textbf{#2}}
\newcommand*\bfcaption[2]{\caption[#1]{\textbf{#1.}#2}}
\usepackage{xcolor}
\definecolor{UBcolor}{HTML}{007CC1}
\usepackage{xurl}

\newcommand{\Movie}[1]{\hyperref[movies]{Supplementary\ Movie\ {#1}}}
\newcommand{\Movies}[2]{\hyperref[movies]{Supplementary\ Movies\ #1}\ and\ \hyperref[movies]{#2}}
\newcommand{\EDFig}[1]{\hyperref[#1]{Supplementary\ \cref{#1}}}
\newcommand{\TheoreticalModel}[1]{\hyperref[#1]{Dynamic Model\ \ref{#1}}}

\newcommand{\bu}{\mathbf{u}}

\usepackage[colorlinks=true,pdfnewwindow=true,linkcolor=UBcolor,citecolor=UBcolor,urlcolor=UBcolor,breaklinks=true,linktocpage]{hyperref}
\usepackage[all]{hypcap}
\usepackage[nameinlink,capitalise]{cleveref}
\crefname{SI section}{SI Section}{SI Sections}
\Crefname{SI section}{SI Section}{SI Sections}

\def \TITLE {Spontaneous Motion of Floating Ice Discs}

\def \TITLE {Hydrodynamic Instability Explains Spontaneous Motion of Floating Ice Discs}

\def \TITLE {Hydrodynamic Instability Induces Spontaneous Motion of Floating Ice Discs}

\def \ie {\textit{i.e.,}~}

\def \lp {\left(}
\def \rp {\right)}
\def \deg {\,^{\circ}\mathrm{C}}

\def \tp {\tilde p}                
\def \T {T}                
\def \tT {\tilde{T}}                

\def \Tc {T_{\mathrm{c}}}            
\def \Th {T_{\mathrm{h}}}             
\def \Tf {T_{\mathrm{f}}}                 
\def \H  {H}                       
\def \WH  {\mathcal{H}}                         
\def \Asy {\mathcal{A}_{\alpha}}

\def \epsiA {\epsilon_{\mathcal{A}}}
\def \U {\mathcal U}                                    
\def \V {\mathcal V}                                       
\def \a {a}                                          
\def \b {b}                                          
\def \g {g}                                         
\def \tt {\tilde t}                                         

\def \Rac {\Rap^*}                                    

\def \Ram {\Rap^{\mathrm{max}}}               

\def \Pr {\mathrm Pr}
\def \lsys {\mathcal{L}}                                     
\def \Nusys {\mathrm{Nu}}                                   
\def \Tw {T_{\rm w}}                                          
\def \be {\mathbf{e}}
\def \mT {\overline T}                                      

\def \tu {\tilde  {\mathbf   u}}                                

\def \zdir {{\mathbf   e}_z}

\def \Ras {\mathrm{Ra}} 

\def \Rap {\mathrm{Ra}_{\mathrm{p}}} 

\def \alphag {\alpha_{\rm gt}}

\def \Trdim {T_{\mathrm{TMD}}}
\def \Trnon {\mathcal{T}}

\def \mS {\mathcal{S}}

\begin{document}

\title{\TITLE}

\author{Min Li}
\affiliation{Department of Mechanical Engineering,  National University of Singapore, 117575, Singapore}

\author{Lailai Zhu}
\email{lailai\_zhu@nus.edu.sg}
\affiliation{Department of Mechanical Engineering,  National University of Singapore, 117575, Singapore}

\date{\today}

\begin{abstract}

Spinning ice discs in nature have been reported for more than a century, yet laboratory experiments have yielded diverse observations and contradictory  explanations, leaving the mechanism behind the disc motion elusive. Here we combine numerical simulations and 
scaling analysis
to investigate a freely-moving ice disc in a lab-scale water tank. We observe the disc remaining stationary or experiencing spontaneous motion, depending on the disc-water temperature difference and water depth. The motion is initiated by a buoyancy-driven, downward plume 
arising from water’s density anomaly---its density peaks near $4\deg$. 
Crucially, the plume breaks rotational and mirror symmetries after descending beyond a critical distance due to a thermoconvective instability, 
thereby inducing the disc to move autonomously. 
Our findings quantitatively unify disc behaviors observed across independent experiments and 
establish a predictive criterion for the onset of disc motion.
More broadly, we point to a route for thermally-driven transport: coupling of bulk thermoconvection and moving bodies, relevant to geophysical processes such as continental drift and iceberg capsizing.

\end{abstract}

\maketitle

\begin{figure*}[tbhp!]
\begin{center}
\includegraphics[scale=1.0]{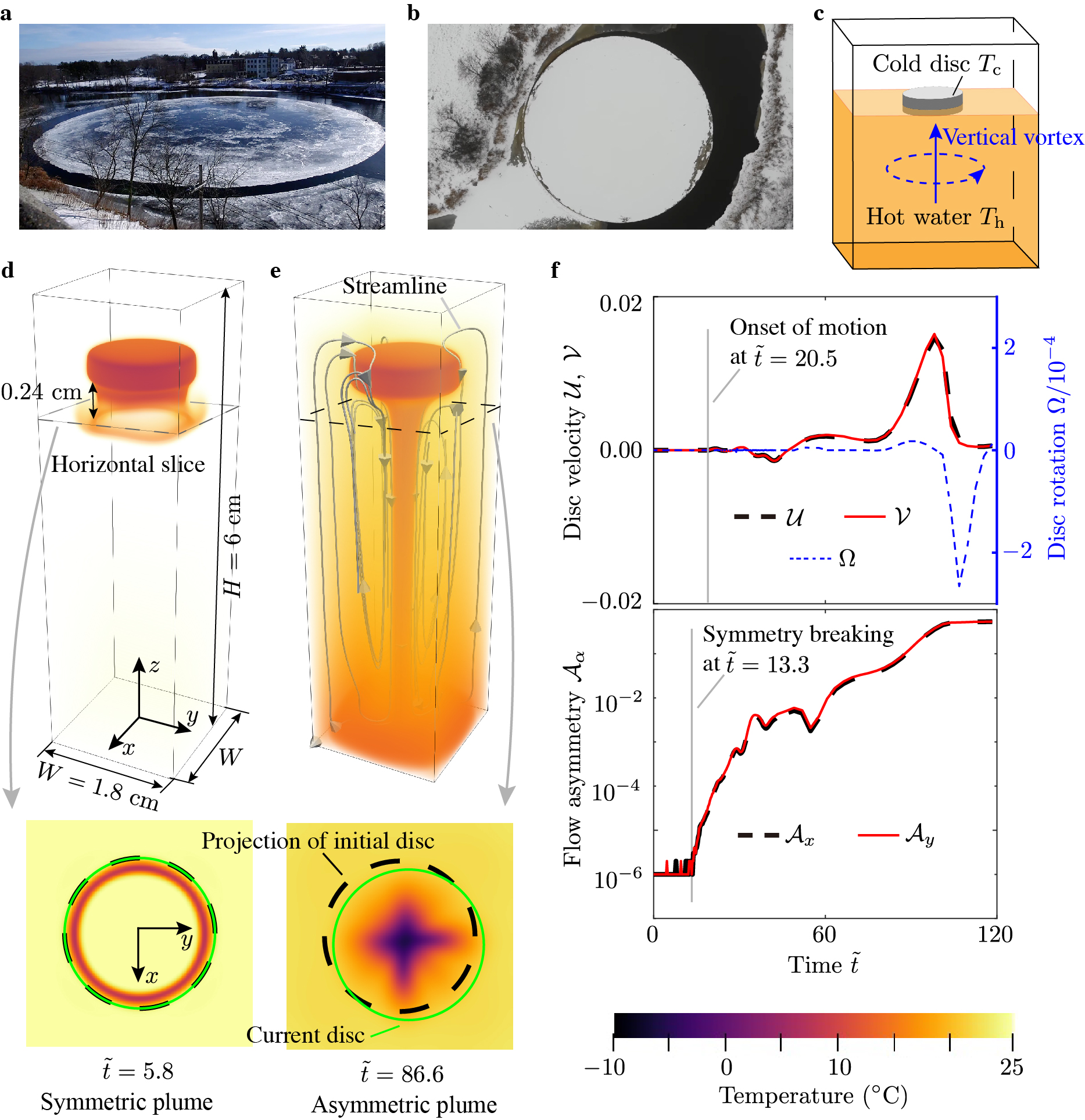}
\end{center}
  {\phantomsubcaption\label{Fig field disc1}}
  {\phantomsubcaption\label{Fig field disc2}}
  {\phantomsubcaption\label{Fig experimental disc}}
  {\phantomsubcaption\label{Fig neumerial disc early}}
    {\phantomsubcaption\label{Fig neumerial disc later}}
  {\phantomsubcaption\label{Fig neumerial disc 2}}
\bfcaption{Simulations reproduce spontaneous motion of an ice disc}
{
\subref*{Fig field disc1} and \subref*{Fig field disc2}, Field observations of rotating ice discs on the Presumpscot River (2019; diameter $\approx 91$ m)~\citep{bangor2019ice} and Vigala river (2019; diameter over 20 m)~\citep{jeeser2019ice}, respectively. 
\subref*{Fig experimental disc},  Schematic of early
laboratory experiments  with centimeter-scale ice or plastic discs floating on quiescent water~\citep{dorbolo2016rotation,schellenberg2023rotation, kistovich2024self,chaplina2024vortex}.  
The arrowed circle indicates a vertical vortex hypothesized by \citet{dorbolo2016rotation}.
We simulate a chilled, non-melting disc immersed in warm water within a closed tank, as detailed in the main text. Initially ($\tt=5.8$), the temperature field is axisymmetric, and the disc is stationary and centered (\subref*{Fig neumerial disc early}); by $\tt=86.6$, the field is asymmetric and the disc has shifted off-center spontaneously (\subref*{Fig neumerial disc later}). 
\subref*{Fig neumerial disc 2}, Temporal evolution of the disc velocity and flow asymmetry. 
}
\label{Fig moving discs}
\end{figure*}

The fluid dance of rotation shapes nature profoundly, from the violent spirals of tornadoes and waterspouts to the grand, slow swirls of oceanic gyres. In a captivating counterpoint, water performs this same dance even in its solid state, giving rise to the rare and enigmatic spectacle of rotating ice discs. 
First documented in 1895~\cite{bates1895revolving}, these natural rarities---large, circular discs of ice, typically 1 to 100 metres in diameter, gently spin within melt holes in partially frozen rivers (see \cref{Fig field disc1,Fig field disc2}).
Such ice discs emerge sporadically in cold regions across North America \citep{gibbens2019howmainesgiantspinningicediskformed,pritchard2022spinningicedisk}, Scandinavia \citep{nordell1997large,thelocal2015swedenicecircle}, Russia~\citep{russia2016spinning,weather2025icecircle}, United Kingdom~\citep{telegraph2009_iceDiscUK,davies2023incredibly}, and Northern China \citep{cgtn2018spinning_ice_circle_NE_China,jiang2020giant}. 
Thought to originate from irregular ice floes sculpted by contact abrasion with surrounding ice \citep{nordell1997large}, these discs can maintain clockwise or counterclockwise rotation at a rate of approximately $5\times 10^{-3}-2\times 10^{-2}\,\text{rad}\,\text{s}^{-1}$
for weeks to months  \citep{nordell1997large}. 
Besides, they occasionally drift from surrounding ice \citep{nordell1997large}, indicating their translational motion in addition to rotation.

Despite over a century of observations, the motion 
of ice discs defies explanation, hindered by their sporadic occurrence in remote, harsh environments. This knowledge gap has spurred laboratory experiments, pioneered by \citet{dorbolo2016rotation}, to demystify this phenomenon.
Although these studies all examined centimetre-scale ice or plastic discs floating on quiescent water 
(\cref{Fig experimental disc})~\citep{dorbolo2016rotation,schellenberg2023rotation,kistovich2024self,chaplina2024vortex}, their divergent observations and explanations have deepened the debate over the underlying mechanism.
\citet{dorbolo2016rotation} attributed their observed translation and rotation of discs to water's density anomaly---where density peaks 
at $\approx 4 \deg$, known as the temperature of maximum density (TMD). 
As the ice cools surrounding water to TMD, denser cold fluid sinks, forming a downward plume that further induces a vertical vortex to drive the disc (\cref{Fig experimental disc}). However, \citet{schellenberg2023rotation} argued that this vortex cannot produce disc translation, and importantly, their flow visualizations did not detect such vortices. They proposed instead that residual flows from tank preparation trigger spontaneous motion. 
This explanation is nevertheless contradicted by their own experiments: ambient-temperature plastic discs remained stationary despite subject to residual flows. Moreover, their observations of chilled counterparts moving akin to ice discs in fact partially corroborate the sinking-plume mechanism suggested by \citet{dorbolo2016rotation}. 
This motion of plastic discs also challenges recent claims that non-uniform melting of ice initiates disc rotation~\citep{kistovich2024self,chaplina2024vortex}.

Here, we revisit this mysterious phenomenon from a computational/theoretical hydrodynamics perspective. Using direct numerical simulations,  we investigate thermally-driven flows around a freely-moving ice disc in quiescent water.  Through simulations devoid of externally imposed disturbances, we observe a chilled disc remaining stationary or undergoing spontaneous translation/rotation---dictated by thermal-hydrodynamic conditions.
We reveal that the  disc's spontaneous motion is initiated from a buoyancy-driven hydrodynamic instability due to  water's density anomaly, and, crucially, the subsequent nonlinear interaction of unstable thermal-convective patterns.

\section*{Results and Discussion} \label{results}

\subsection{Spontaneous Motion in an Ideal Configuration} 
\label{ideal configruation}
To isolate the mechanism, we consider a minimal numerical setup  of a floating ice disc, neglecting air-disc interactions, heat conduction within the disc, and  melting, as justified in \cref{assumption}.  
Specifically, we simulate an isothermal, non-meltable  rigid disc of radius $\a=0.6\,$cm and thickness $\b=0.12\,$cm,  immersed in a sealed,  water-filled,  square-based cuboidal tank  (width $W=1.8\,$cm, height $H=6 \, $cm), with insulated boundaries, see \cref{Fig neumerial disc early}.  Initially, the water is quiescent and uniformly set to temperature $\Th=25 \deg$. 
The disc, maintained at $\Tc = -10\deg$, is  initially at rest and centered in the tank,  $0.9\,$cm  below its upper wall. 
It can freely translate in the horizontal ($xy$) plane and rotate about the vertical ($z$) axis, while constrained vertically. 
This setup excludes externally imposed disturbances, such as residual background flow or non-uniform ice melting.

Our numerical simulations  reproduce the spontaneous motion of disc, which is found to be coupled with the spatiotemporal evolution of the water temperature. 
Specifically, at the early stage ($\tt = 5.8$, \cref{Fig neumerial disc early}), the disc remains stationary and centered, while the surrounding water cooled to TMD ($\Trdim = 4 \deg$), sinks and forms an annular, skirt-like plume.
This structure appears axisymmetric, as demonstrated by  the temperature field in a plane 0.24 cm below the disc bottom (\cref{Fig neumerial disc early}). 
By $\tt=86.6$ (\cref{Fig neumerial disc later}), the developing plume becomes asymmetric and meanwhile the disc spontaneously moves off-center, implying a causal
relationship between the two processes. Before confirming it, we assert that this spontaneous motion cannot be explained by the hypotheses of residual background flows~\citep{schellenberg2023rotation} or non-uniform melting~\citep{kistovich2024self,chaplina2024vortex}, as both are excluded numerically by design.
Furthermore, vertical vortices as speculated~\cite{dorbolo2016rotation} 
are not observed here (see \cref{Fig neumerial disc later}), aligning with the experimental findings~\citep{schellenberg2023rotation}; in fact, we identify a ring of azimuthal vortices, characterized by  closed streamlines in the vertical plane (see \cref{Fig neumerial disc later}).

To examine the connection between the plume asymmetry and disc motion, 
we quantify the flow's mirror asymmetries $\mathcal{A}_{x}$ and $\mathcal{A}_{y}$ (see \cref{criterion}) across the vertical midplanes  $x=0$ and $y=0$, respectively. 
We track their temporal evolutions in \cref{Fig neumerial disc 2}, 
alongside those of disc's dimensionless horizontal velocity  $\U\be_x + \V\be_y$ and rotational velocity  $\Omega\be_z$. 
As illustrated, the flow loses symmetry at the dimensionless time $\tt \approx 13.3$, whereas the disc starts moving later, at $\tt \approx 20.5$.  This lag is expected, since the asymmetric flow requires time to accelerate the disc. 
While the disc is driven by the asymmetric plume flow,
we uncover that the asymmetry results from a buoyancy-driven hydrodynamic instability, where the evolution of 
intrinsic (naturally occurring) perturbations, not extrinsic disturbances,  plays a dominant role.

\begin{figure*}[tbhp!]
\begin{center}
\includegraphics[scale=1.0]{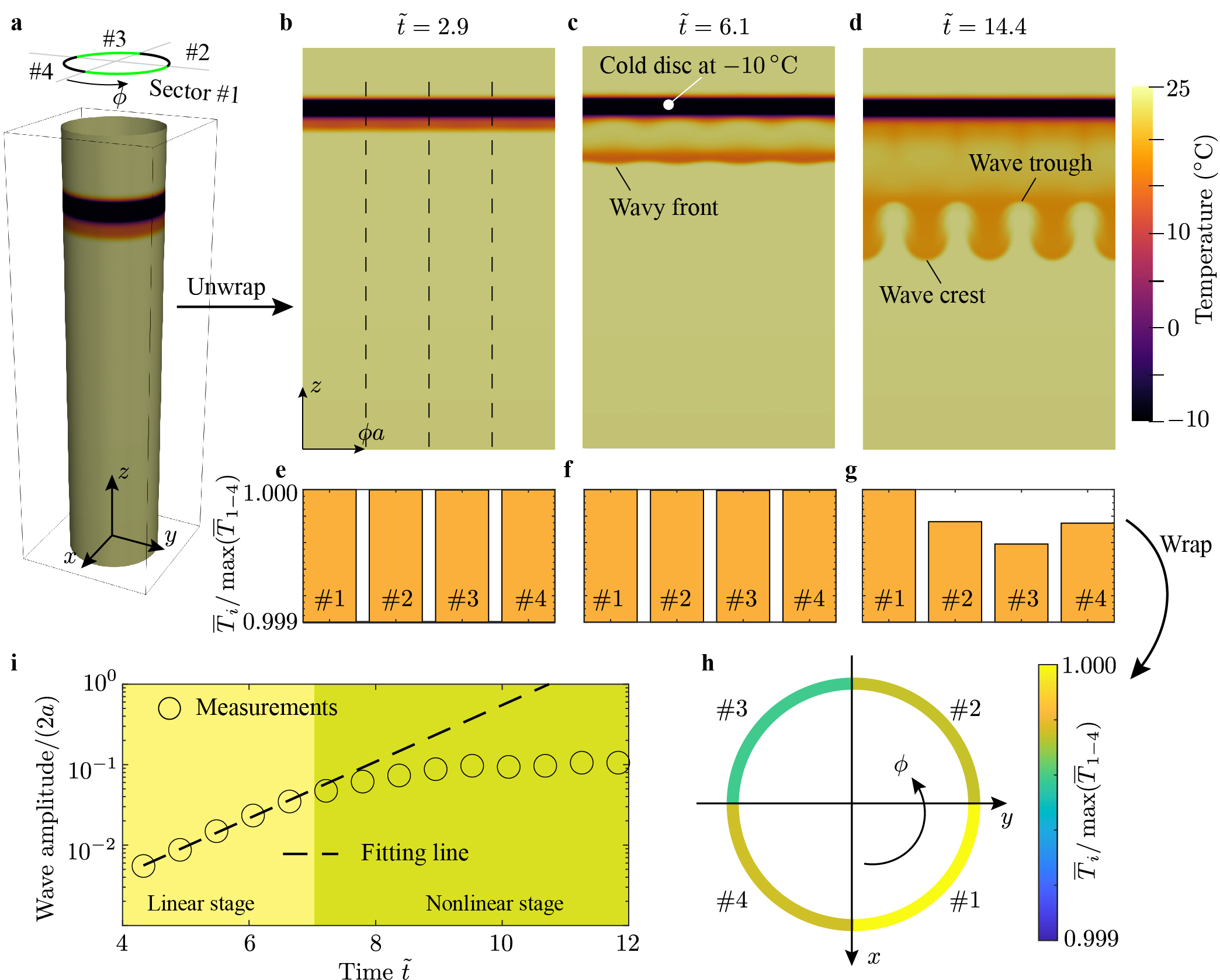}
\end{center}
  {\phantomsubcaption\label{temperature slice}}
    {\phantomsubcaption\label{unwraped slice 1}}
    {\phantomsubcaption\label{unwraped slice 2}}
    {\phantomsubcaption\label{unwraped slice 3}}
    {\phantomsubcaption\label{Ti 1}}
     {\phantomsubcaption\label{Ti 2}}
 {\phantomsubcaption\label{Ti 3}}
     {\phantomsubcaption\label{wrapped Ti 3}}
  {\phantomsubcaption\label{LSA growth}}

  \bfcaption{Thermoconvective flow breaks symmetry spontaneously} 
  { 
\subref*{temperature slice}, Temperature distribution on a cylindrical slice extended from the disc's lateral side. The slice is divided into four equal azimuthal ($\phi$) sectors. 
\subref*{unwraped slice 1}--\subref*{unwraped slice 3}, Unwrapped temperature distribution over times, highlighting the progressive undulations of the plume front. 
\subref*{Ti 1}--\subref*{Ti 3}, Azimuthal sector-mean temperatures at successive times, 
corresponding to \subref*{unwraped slice 1}--\subref*{unwraped slice 3}, respectively.  
\subref*{wrapped Ti 3}, Wrapped representation of \subref*{Ti 3}, 
showing broken rotational and mirror symmetries.  
\subref*{LSA growth}, Temporal evolution of the amplitude of the plume front.  
}

\label{symmetry breaking}
\end{figure*}

We characterize this instability by scrutinizing the spatiotemporal temperature field on an unwrapped cylindrical slice attached to the disc's sidewall, see \cref{temperature slice}. Upon partitioning this slice into four equal azimuthal ($\phi$) sectors, we monitor the area-weighted mean temperature $\mT$ over each, denoted $\mT_i$ for the $i$-th sector. 
The evolution of temperature exhibits signatures of Rayleigh-Taylor (RT)
instability---a paradigmatic instability that occurs when a denser fluid lies atop a lighter one with both under gravity  \citep{SHARP19843,suchandra2023dynamics}. 

At the early stage $\tt=2.9$ in \cref{unwraped slice 1}: a belt of colder, denser fluid beneath the ice disc  forms above the remaining bulk of  warmer, lighter fluid. The former sinks and intrudes into the latter, generating a downward-propagating flat front. This pattern is effectively the planform manifestation of the plume on the cylindrical slice. 
By  $\tt=6.1$, it develops a four-fold azimuthal wavy pattern  (see \cref{unwraped slice 2})  that becomes increasingly pronounced by $\tt=14.4$ (see  \cref{unwraped slice 3}), characteristic of RT instability.
The wave amplitude of the front (defined in  \cref{Front Amplitude}),   grows exponentially until   $\tt\approx 7$ and subexponentially thereafter    (see \cref{LSA growth}), reflecting the transition from the linear to the nonlinear stage of  instability. 
During the linear stage  ($\tt<7$), the instability-induced perturbations are weak  and thus the sector-mean temperature  $\mT_i$ stays  azimuthally uniform (see  $\tt=2.9$ and $\tt = 6.1$, \cref{Ti 1,Ti 2}). 
In the nonlinear stage  ($\tt>7$), $\mT_i$ develops azimuthal nonuniformity across the four sectors  (see  $\tt=14.4$,  \cref{Ti 3}),  due to nonlinear interactions among perturbations (wave crests/troughs)   \citep{SHARP19843,suchandra2023dynamics}. 
Mapping  $\mT_i$ onto a circle
(\cref{wrapped Ti 3})  shows that the plume flow  has lost both rotational and mirror symmetries, thereby enabling the disc motion. In summary, spontaneous disc motion arises from the RT instability, 
which breaks the symmetries of flow through nonlinear interactions of perturbations, leading to a hydrodynamic forcing on the disc.

\begin{figure*}[tbhp!]
\begin{center}
\includegraphics[scale=1.0]{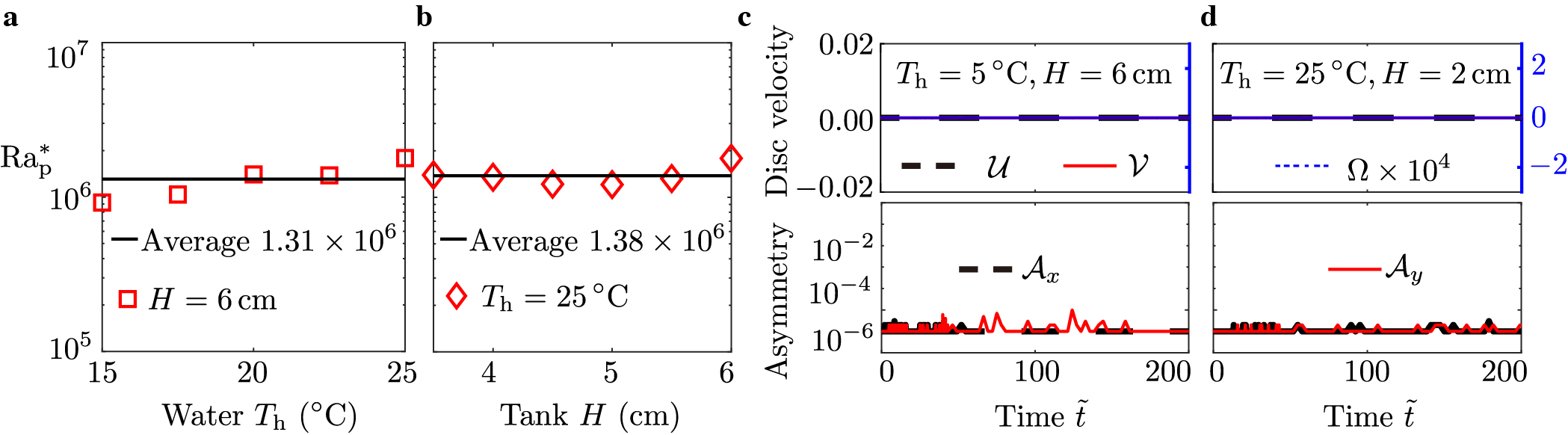}
\end{center}
  {\phantomsubcaption\label{Rac at various Th}}
   {\phantomsubcaption\label{Rac at various H}}
  {\phantomsubcaption\label{Numerical results below Rac cold water}}
  {\phantomsubcaption\label{Numerical results below Rac short tank}}
\bfcaption{
Criterion for the onset of disc motion and symmetry breaking}{
\subref*{Rac at various Th} and \subref*{Rac at various H}, Critical plume Rayleigh number $\Rac$ versus the initial water temperature $\Th$ at fixed tank height $\H = 6$ cm, and $\Rac$ versus $\H$ at $\Th = 25 \deg$, respectively. 
\subref*{Numerical results below Rac cold water}, Temporal evolution of disc velocity and flow asymmetry for $\Th=5\deg$ instead of  the baseline $\Th=10\deg$.
\subref*{Numerical results below Rac short tank},  As in \subref*{Numerical results below Rac cold water}, but with the baseline $\H=6$ cm reduced to $\H = 2$ cm.
}
\label{Fig Transition Rac}
\end{figure*}

Having traced the spontaneous disc motion to the nonlinear evolution of an RT instability, we proceed to establish a criterion for the onset of motion. This is equivalent to determining when the flow loses symmetry. Following prior works on symmetry breaking of smoke plumes~\citep{kimura1983mechanism, noto1989swaying, yang1992buckling}, we define the plume Rayleigh number 
\begin{equation} \label{streamwise Ra}
\Rap = \frac{\g \alphag (\Th-\Trdim)^{1.895} h^3}{\nu \kappa},
\end{equation}
where $h$ denotes the instantaneous plume length, measured from the disc bottom to the plume front (see \cref{Front Amplitude}). 
This number compares the buoyancy forcing across the length $h$ against viscous and thermal diffusion; 
here, $\g$ is the gravitational acceleration, $\alphag$ the generalized thermal expansion coefficient,  $\nu$  the kinematic viscosity, and  $\kappa $ the thermal diffusivity, as detailed in \cref{sec:equations}. 
We then numerically pinpoint the critical plume Rayleigh number $\Rac$ for flow symmetry breaking and concomitant disc motion. By varying the initial water temperature $\Th$ at a fixed tank height  $\H=6$ cm (\cref{Rac at various Th}) or changing $\H$ while holding  $\Th =25 \deg$ (\cref{Rac at various H}), we identify $\Rac \approx 1.3 \times 10^6$ that is insensitive to height $\H$ and  temperature $\Th$. Moreover, this threshold depends weakly on the tank geometry,  boundary conditions, and disc temperature, as evidenced in \cref{Critical Ra in SI}.

Because $\Rap$ encodes the plume length $h$, its upper bound $\Ram$ occurs when the plume attains its maximum length  $\WH$,  corresponding to the depth of water beneath the disc.
Comparing $\Ram$ with the critical value $\Rac$ yields the criterion for triggering the disc motion and symmetry breaking, 
\begin{equation} \label{condition of symmetry breaking}
\Ram =\frac{\g \alphag (\Th-\Trdim)^{1.895} \WH^3}{\nu \kappa}  \gtrsim \Rac  \approx 1.3 \times 10^6.
\end{equation}
This condition holds for the illustrative baseline
setup in \cref{Fig moving discs}, where $\Th=25~\deg$ and $\H=6 $ cm give $\Ram=2.6\times 10^7$. Hence, the disc spontaneous moves, and the flow becomes asymmetric (\cref{Fig neumerial disc 2}).  
Relative to the baseline, lowering either the water temperature to $\Th=5~\deg$ where $\Ram =8.1\times 10^4 <\Rac$ (\cref{Numerical results below Rac cold water}) or the tank height to $\H = 2$ cm where $\Ram=2.6\times 10^5 <\Rac$ (\cref{Numerical results below Rac short tank}) yields a stationary disc without symmetry breaking; 
both tests support the identified criterion \cref{condition of symmetry breaking}.

Furthermore, this condition offers a unified explanation for the experimental observations of \citet{dorbolo2016rotation} and \citet{kistovich2024self}. First, the positive dependence of $\Ram$ on water temperature $\Th$ and water depth $\WH$ rationalizes the general trend reported by both experiments \citep{dorbolo2016rotation,kistovich2024self}: the disc spontaneously moves in hot, deep water, but remains stationary in either cold or shallow water. 
Second, \cref{condition of symmetry breaking} consistently captures two experimentally identified thresholds for initiating disc motion: (i) $\Th \geq 5 \deg$ at  $\WH = 12.5$ cm~\citep{dorbolo2016rotation} and (ii) $\WH \geq 3$ cm when $\Th = 20 \deg$~\citep{dorbolo2016rotation,kistovich2024self}. These map to 
$\Ram \geq 1.2 \times 10^6$ and   $\Ram \geq3.1 \times 10^6$,  respectively, matching our criterion \cref{condition of symmetry breaking} within reasonable experimental variability.

\subsection{Can External Disturbances Alone Drive the Disc Motion?} 
\label{Practical configruation}

The disc motion  has so far been examined in the absence of  external forcing,  and arises solely from the growth of intrinsic perturbations  associated with RT instability.
We now investigate whether externally imposed disturbances alone, such as residual  flows, can drive the disc motion as hypothesized~\citep{schellenberg2023rotation}. 

\begin{figure*}[tbhp!]
\begin{center}
\includegraphics[scale=1.0]{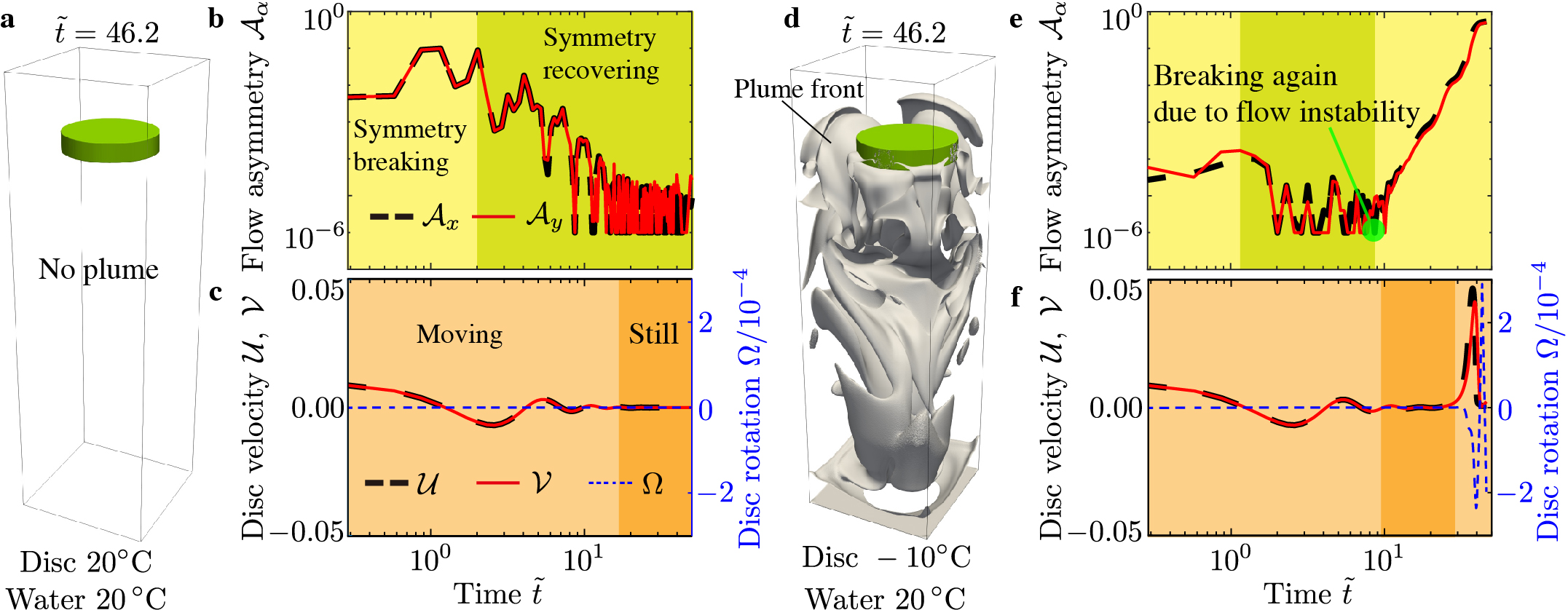}
\end{center}
  {\phantomsubcaption\label{Fig Th=Tc plume}}
  {\phantomsubcaption\label{Fig Th=Tc asymmetry}}
   {\phantomsubcaption\label{Fig Th=Tc disc}}
  {\phantomsubcaption\label{Fig Th>Tc plume}}
  {\phantomsubcaption\label{Fig Th>Tc asymmetry}}
   {\phantomsubcaption\label{Fig Th>Tc disc}}
\bfcaption{
Residual flow does not significantly alter instability-driven disc motion
}  
{
\subref*{Fig Th=Tc plume}--\subref*{Fig Th=Tc disc},
Residual flows in an isothermal setting ($\Th = \Tc =20 \deg$; $\H =6$ cm) induces transient, short-lived flow asymmetry and disc movement. 
\subref*{Fig Th>Tc plume}--\subref*{Fig Th>Tc disc},  In contrast, with $\Th=20>\Tc=-10 \deg$, a temperature difference triggers RT instability, which---after the residual-flow transient---persistently drives flow asymmetry and disc motion. 
}
\label{Fig Perturbation effects}
\end{figure*}

Specifically, we adjust the above-studied baseline case by setting the disc temperature to the ambient water temperature, \ie $\Tc=\Th=20\deg$, corresponding to an isothermal configuration. 
To mimic the residual flows,  we initialize the flow with a random velocity disturbance everywhere with magnitude $0.1-0.5$ mm/s, consistent with the experimental conditions~\citep{schellenberg2023rotation}. 
\cref{Fig Th=Tc plume} shows  
the absence of buoyancy-driven plumes, as anticipated in the isothermal setting.
Meanwhile, the random impulsive forcing induces flow asymmetry, which experiences a transient amplification and then  decays to a negligible level (see \cref{Fig Th=Tc asymmetry}). Accordingly, the disc undergoes a temporary oscillation and then recovers to a stationary state (see \cref{Fig Th=Tc disc}). The recovery is expected because the initial disturbances are dampened by viscous dissipation. 
This explains the experimental observation of \citet{schellenberg2023rotation} that plastic disks at ambient water temperature  ($\Tc=\Th$) end up in motionless.

We now activate thermal effects by setting $\Tc=-10\deg$ under otherwise identical conditions, which then produces a plume (see \cref{Fig Th>Tc plume}). Early on ($\tt \lessapprox 10$), the flow and disc respond as in the isothermal scenario---both asymmetry and motion gradually vanish (see \cref{Fig Th>Tc asymmetry,Fig Th>Tc disc}). 
Thereafter, the flow regains asymmetry at $\tt \approx 10$, and the disc restarts moving  at $\tt \approx 30$ as shown in \cref{Fig Th>Tc asymmetry,Fig Th>Tc disc}---this time caused by the nonlinear interactions of intrinsic perturbations.

We conclude that external impulsive disturbances alone can generate short-lived disc motion. 
However, these disturbances gradually die out due to viscous dissipation and thus  cannot sustain the motion. 
This reinforces the RT  instability as the key mechanism underlying experimentally observed disc motion~\citep{dorbolo2016rotation,schellenberg2023rotation,kistovich2024self,chaplina2024vortex}.

\section*{Conclusions and Discussions}  \label{conclusion}

We reveal a buoyancy-driven pathway for the spontaneous motion of ice discs in water: 
thermal convection coupled to macroscopic object motion. A cold disc initiates a descending thermal plume by exploiting water's density anomaly; subsequently---and critically---nonlinear interactions among instability-induced perturbations break flow symmetries beyond a threshold plume length, inducing a net  hydrodynamic forcing that drives the disc.

A comparable scenario is the emergent oscillation of a plate floating on the free surface of a turbulent
Rayleigh-B\'enard convective fluid~\cite{zhang2000periodic,zhong2005thermal,mac2018stochastic,mao2019dynamics,mao2021insulating,lowenstein2024near,mac2024covering,mac2025covering}, with possible relevance to continental drift. 
Another relevant observation is the overturn of floating horizontal ice cylinders, which can rotate freely about their  axes---as investigated experimentally to elucidate capsizing icebergs~\citep{johnson2024lab,bellincioni2025melting,johnson2025shape}. 
While similar to our study, their capsize incidents result from a gravitational instability caused by melting-induced shape evolution of ice, whereas our mechanism is a flow instability that involves no melting  or reshaping. 

Collectively, these phenomena, reported both previously  and in this study,  exemplify a feedback-mediated thermoconvective fluid-object 
interaction: the object seeds or modulates the flow, and the flow, in turn, affects the former's motion.

\bigskip

\noindent\textbf{Acknowledgments}
 We  thank  St\'ephane Dorbolo for sharing part of the experimental data. Computation of the work was performed on resources of the National Supercomputing Centre, Singapore (https://www.nscc.sg), as well as those provided by NUS IT via a grant (NUSREC-HPC-00001).

\bigskip

\noindent\textbf{Author contributions}

M.L. performed the simulations and analyses. L.Z. conceived and supervised the research.  Both authors analyzed the data and wrote the manuscript.

\bigskip

\noindent\textbf{Competing interests}

The authors declare no competing interests.

\bigskip

\noindent\textbf{Data and code availability}

The data supporting the findings of this study are available from the corresponding author, Lailai Zhu, upon reasonable request.

\bibliography{ref}

\onecolumngrid
\clearpage
\setcounter{figure}{0}
\renewcommand{\figurename}{Supplementary Figure}
\renewcommand{\thefigure}{S\arabic{figure}}
\twocolumngrid

\section*{Methods} \label{methods}

\subsection{Assumptions}  \label{assumption}
This study adopts a minimal configuration that neglects  air-disc interactions, heat conduction within the disc, and disc melting. Furthermore, we assume that the disc is fully submerged in water with no vertical motion. We justify these assumptions below.

\subsubsection*{Effects of Air on Ice Disc}
The  density and viscosity of air are significantly lower than those of water. Thus, aerodynamic forces and torques acting on the disc---both inertial and viscous---are negligible compared to their hydrodynamic counterparts. This justifies neglecting the air phase in simulations.

\subsubsection*{Heat Conduction}
Spontaneous motion has been experimentally observed across a range of disc-water temperature differences \citep{dorbolo2016rotation}, indicating that the rate of heat conduction within the ice disc is not essential for initiating the motion. 
Thus, in our simulations, the disc is modeled as isothermal, with the rate of heat conduction explicitly set to zero.

\subsubsection*{Ice Melting}
The experiments by \citet{schellenberg2023rotation} demonstrated that both ice and plastic discs can spontaneously move under identical conditions, thereby ruling out ice melting as a necessary condition for spontaneous motion. Consequently, we model the disc as a non-melting rigid object.

\subsubsection*{Submerged and Vertically Constrained Disc}
The ice disc exposes only $\approx 10\%$ of its volume to air \citep{cenedese2023icebergs,bellincioni2025melting,sweetman2025influence} 
and exhibits negligible vertical motion \citep{baskin2020heat}. Therefore, it is reasonable to assume a fully submerged disc without vertical displacement.

\subsubsection*{\textit{A Posteriori} Verification}
As demonstrated in the main text, our simulations successfully reproduce various experimental observations, justifying  the invoked assumptions \textit{a posteriori}.

\subsection{Governing Equations} 
\label{sec:equations}
We consider a thermally-driven  flow, \ie how the velocity $\bu$, pressure $p$, and temperature $T$ of fluid involves in time $t$. 
Assuming negligible variations of all fluid properties with the temperature except for the 
density change in buoyancy 
\citep{wang2021ice,wang2021equilibrium,wang2021growth,weady2022anomalous,xia2023tuning}, the dimensional governing equations are 
\begin{subequations} \label{dimensional governing equations}
\begin{align}
\grad \cdot \bu &= 0,\\
\frac{\partial   \bu}{\partial t} +   \bu \cdot \grad   \bu &= g\alphag |T-\Trdim|^{1.895}\zdir + \nu \nabla^2  \bu - \frac{\grad  p}{\rho_0},\\
\frac{\partial T}{\partial t} + \bu \cdot \grad T &=  \kappa \nabla^2 T,
\end{align}
\end{subequations}
where $\rho_0$, $\nu$, and $\kappa$ denote the reference density, kinematic viscosity, and thermal diffusivity of the fluid (water here), respectively;
$\alphag = 9.3 \times 10 ^{-6}~{\rm \deg}^{-1.895}$ is the generalized thermal expansion coefficient accounting for the density anomaly  \citep{gebhart1977new,yang2022abrupt,wang2021growth,du2024physics,johnson2025shape}; $\g=9.8~\rm{m/s^2}$ is the gravitational acceleration; $\zdir$ represents the upward basis unit vector. Moreover, the  translational and  rotational velocities of the disc are determined by Newton’s second law. 
Here, we set $\rho_0 =  999.972~{\rm kg/m^3}$, $\nu = 10^{-6} ~{\rm m^2/s}$, and $\kappa = 1.5\times 10^{-7}~{\rm m^2/s}$.

We nondimensionalize \cref{dimensional governing equations} by selecting disc diameter $2\a$ as the characteristic length,  $\sqrt{g\alphag(\Th-\Tf)^{1.895}2\a}$, $\Th-\Tf$, and $\rho_0 g\alphag(\Th-\Tf)^{1.895}2\a$ as the characteristic velocity, temperature, and pressure, respectively. Here, $\Tf = 0\deg$  denotes the fusion point of ice. 
Using the tilde symbol to represent dimensionless variables, we obtain the dimensionless equations:
\begin{subequations} \label{governing equations}
\begin{align}
\grad \cdot \tu &= 0,\\
\frac{\partial \tu}{\partial \tt} + \tu \cdot \grad \tu &= |\tT-\Trnon|^{1.895}\zdir + \sqrt{\frac{\Pr}{\Ras}} \nabla^2\tu - \grad \tp,\\
\frac{\partial \tT}{\partial \tt} + \tu \cdot \grad \tT &=  \sqrt{\frac{1}{\Ras\Pr}} \nabla^2\tT.
\end{align}
\end{subequations}
Here, the dimensionless parameters include the source Rayleigh number $\Ras$, Prandtl number $\Pr$, and dimensionless TMD of water $\Trnon$, 
\begin{subequations} \label{control numbers}
\begin{align}
\Ras &= \frac{\g\alphag(\Th-\Tf)^{1.895}(2a)^3}{\nu \kappa},\\
\Pr &= \frac{\nu}{\kappa},\\
\Trnon &=   \frac{\Trdim-\Tf}{\Th-\Tf}.
\end{align}
\end{subequations}

\subsection{Numerical Methods and Validations} 
\label{sec:numerical}
We numerically solve the governing equations using a Lattice Boltzmann method \citep{guo2013lattice}. 
An immersed moving boundary scheme  ~\citep{chen2020dirichlet,xia2024particle} is employed to enforce the velocity and temperature boundary conditions at the surface of disc, and to compute the hydrodynamic forces and torques on it.

Our numerical implementation involves three major features: thermally-driven flows, fluid-solid (water-disc) interaction, and the density anomaly of water. The first two are validated against a two-dimensional (2D) benchmark---an isothermally heated cylinder settling in a quiescent fluid  \citep{feng2008inclusion,xia2024particle}. The third feature is validated through three-dimensional (3D) Rayleigh-B\'enard convection of cold water near its density maximum \citep{hu2015rayleigh}. 

\begin{figure}[tbhp!]
\begin{center}
\includegraphics[scale=1.0]{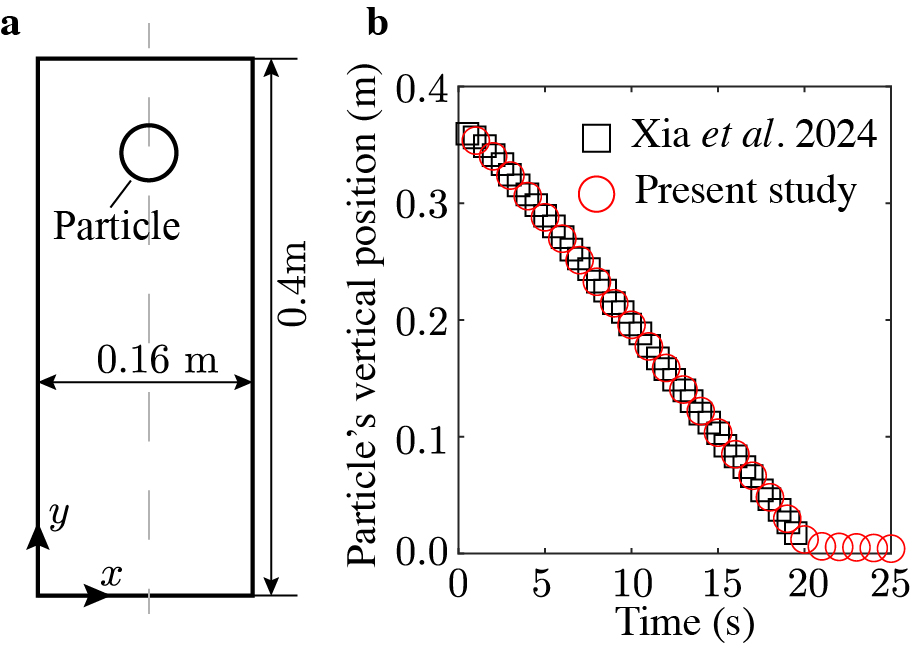}
\end{center}
  {\phantomsubcaption\label{Fig computation domain}}
  {\phantomsubcaption\label{Fig vertical position}}
\bfcaption{Sedimentation of an isothermally heated particle in a quiescent fluid}{ 
\subref*{Fig computation domain}, A 2D rectangular tank filled with quiescent fluid  initially  at 290 K.  A particle, placed on the centerline of the tank and maintained at 300 K, is hotter and heavier than the surrounding fluid. 
\subref*{Fig vertical position}, Temporal evolution of the particle's vertical position from our simulation, benchmarked against \citet{xia2024particle}.
}
  
\label{Fig first benchmark}
\end{figure}

As shown in \cref{Fig computation domain}, the first benchmark adopts a rectangular computational domain of $0.16$ m in width and $0.4$ m in height ($160\times400$ lattice grids), bounded by four no-slip and thermally insulated walls. 
The domain is filled by fluid that is initially quiescent and uniformly at $290$ K.
The fluid has a kinematic viscosity and thermal diffusivity both equal to $10^{-5} ~{\rm m^2/s}$, a density of  1000 $\rm Kg/m^3$, and a thermal expansion coefficient of  $1.02 \times 10^{-4}~ {\rm K}^{-1}$. A cylinder of diameter 0.01 m  is initially positioned  at the horizontal midpoint of the domain and offset vertically from its ceiling by $0.04$ m. 
The cylinder, with a density of 1006.8 $\rm Kg/m^3$, is slightly heavier than the fluid and maintained at 300 K, thus heating the surrounding fluid as it sediments. The temporal evolution of the cylinder’s vertical position is shown \cref{Fig vertical position}, where the excellent agreement between our results and those of \citet{xia2024particle}  
validates our numerical implementation of a moving body in thermally-driven flows.

\begin{figure}[tbhp!]
\begin{center}
\includegraphics[scale=1.0]{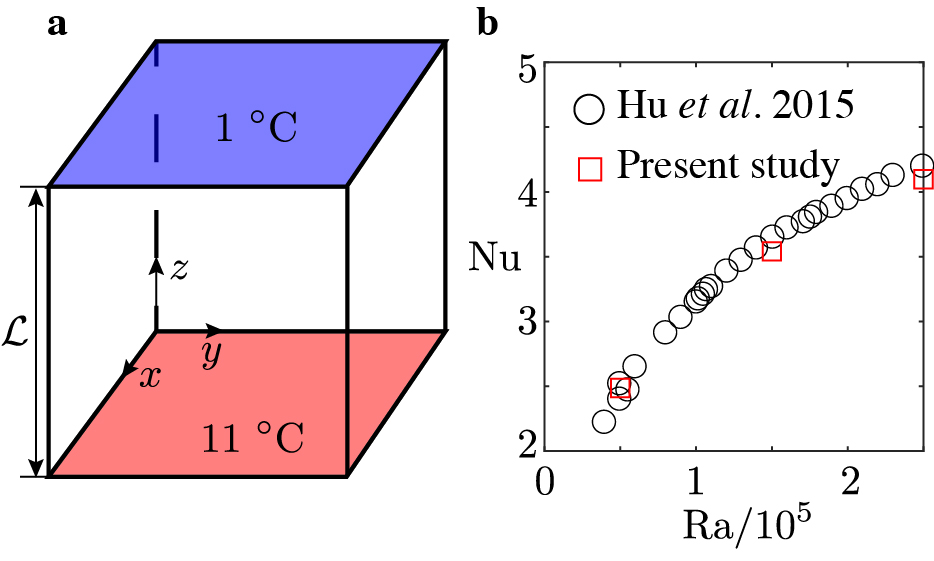}
\end{center}
  {\phantomsubcaption\label{Fig cubical cavity}}
  {\phantomsubcaption\label{Fig Ra_Nu}}
\bfcaption{Rayleigh-B\'enard convection of water 
near TMD}{ 
\subref*{Fig cubical cavity}, A 3D cubical cavity heated from below at $11\deg$ and cooled from above at $1\deg$, with sidewalls thermally insulated. The cold water in the cavity can be heated up to $\Trdim$, near its density maximum, allowing density anomaly to emerge and influence the convective flow.  
\subref*{Fig Ra_Nu}, Nusselt number $\Nusys$ as a function of the source Rayleigh number $\Ras$. Circles and squares denote reference~\citep{hu2015rayleigh} and present numerical results, respectively.
}

\label{Fig second benchmark}
\end{figure}

The second benchmark corresponds to Rayleigh-B\'enard convection within a 3D cubic cavity bounded by no-slip walls, as shown in \cref{Fig cubical cavity}. 
While the sidewalls of cavity are thermally insulated, its bottom and top walls are maintained isothermally at a temperature of $11\deg$ and $1\deg$, respectively. 
The water in the cavity initially at $1\deg$ undergoes temperature changes between  $1\deg$ and $11\deg$, allowing the occurrences of density anomaly at $\Trdim = 4\deg$. 
We examine how source Rayleigh number $\Ras = \g \alphag (11-1)^{1.895} \lsys^3 / (\nu \kappa)$ influences
heat transfer within the cavity, where $\lsys$ denotes the cavity size.  The efficiency of heat transfer is quantified by the surface-averaged, dimensionless temperature gradient  at the bottom wall ($\tilde{z}=0$)---\ie the Nusselt number $\Nusys = - \langle \partial \tT / \partial \tilde z \rangle_{\tilde z=0}$, where $\langle \cdot \rangle_{\tilde z=0}$ denotes a surface average.  

Under a grid resolution of  $200\times 200\times 200$,  we obtain    $\Nusys$  values for $\Ras \leq 2.5\times 10^5$. The results shown in  \cref{Fig Ra_Nu}  agree reasonably well with the data of
\citet{hu2015rayleigh}, hence  verifying that our solver can capture the feature of density anomaly.

With the  key three numerical features validated, we confidently develop  the solver for our target setting---thermoconvective flow around a freely-moving disc in water.

\subsection{Definition of Flow Asymmetry}  \label{criterion}

In the main text, flow asymmetry refers to the mirror asymmetry across the vertical midplane $\alpha=0$ ($\alpha:=x$ or $y$). 
We define it as $\Asy = 1- |{\rm cor}\lp\T,~\T_{\alpha}\rp| + \epsiA$, 
where ${\rm cor}\lp\mS,~\mS_{\alpha}\rp$ calculates
the correlation~\citep{puth2014effective} between a scalar field $\mS$ and its mirror image $\mS_{\alpha}$ across the $\alpha=0$ plane. Here, $\epsiA = 10^{-6}$ facilitates logarithmic presentation in \cref{Fig neumerial disc 2}. 
In the symmetric state, the correlation remains 1, so $\Asy$ stays on a plateau at its baseline value $\epsiA$. 
The first systematic deviation from this plateau indicates the onset of symmetry breaking (see \cref{Fig neumerial disc 2}).

\subsection{Robustness of the Critical Plume Rayleigh Number $\Rac$}
\label{Critical Ra in SI}

In the main text, we identify a critical plume Rayleigh number for flow symmetry breaking and concomitant spontaneous disc motion, $\Rac \approx 1.3 \times 10^6$, which is independent of water temperature and depth, as demonstrated in \cref{Rac at various Th,Rac at various H}. Here, by examining three additional variants  beyond the baseline setup, we show that this critical value is independent of the tank geometry, wall boundary conditions, and disc temperature within our parameter range. In all these setups, the initial water temperature is $\Th = 25\deg$ and the tank height is $\H = 6$ cm.

\begin{figure}[tbhp!]
\begin{center}
\includegraphics[scale=1.0]{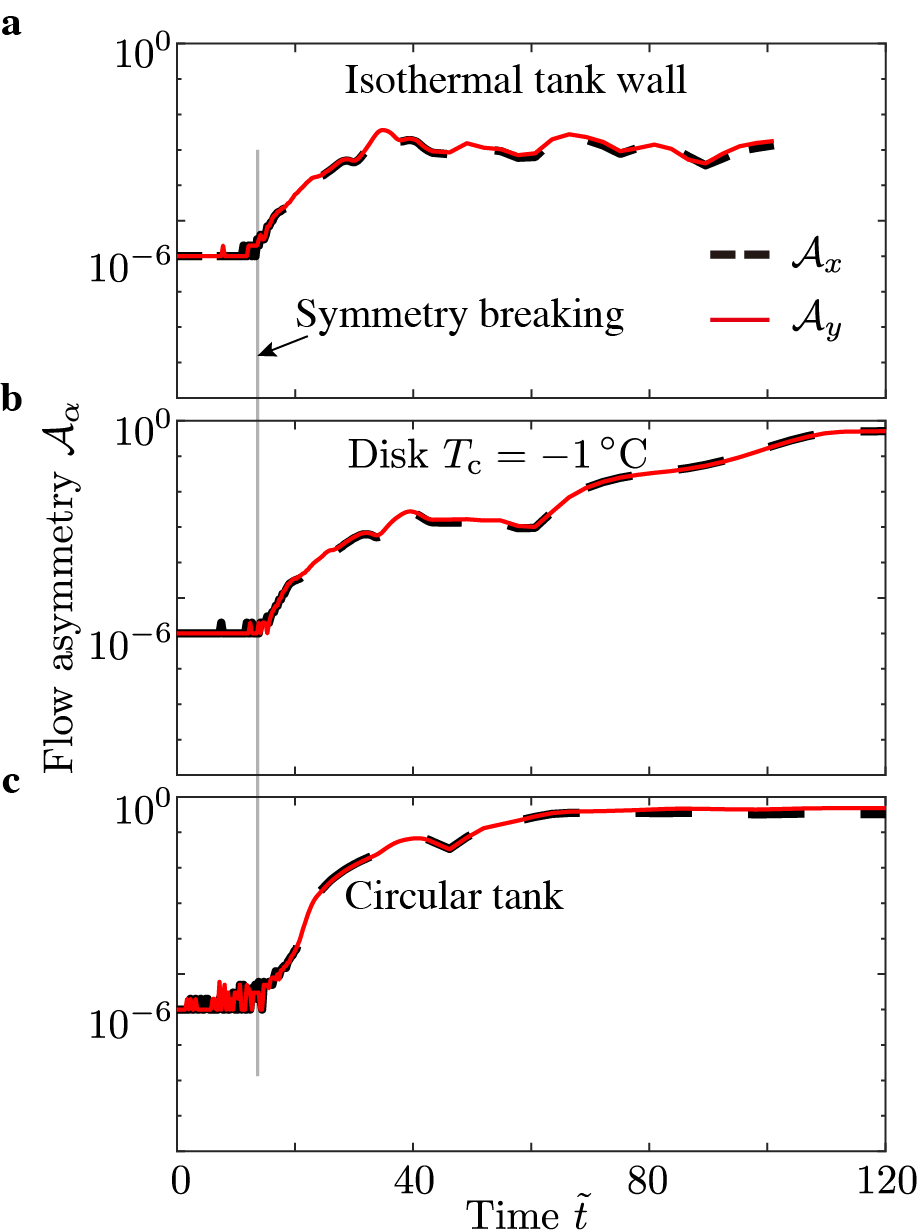}
\end{center}
  {\phantomsubcaption\label{Fig Isothermal tank}}
  {\phantomsubcaption\label{Fig warmer disc}}
 {\phantomsubcaption\label{Fig circular tank}}
\bfcaption{Temporal evolution of flow asymmetry $\Asy$ in three baseline variants}{
Relative to the baseline (\cref{Fig moving discs}), 
we examine: (i) an isothermal enclosure at the water temperature (\subref*{Fig Isothermal tank}); (ii) a warmer disc with $\Tc=-1\deg$ (\subref*{Fig warmer disc}); and (iii) a cylindrical tank replacing the rectangular cuboid (\subref*{Fig circular tank}). 
}
\label{Fig SI_Rac}
\end{figure}

In the first setup, we modify the tank walls from insulated to isothermal, fixing their temperature at $\Tw = \Th$. In the second, the disc temperature is altered, from the baseline value $\Tc = -10\deg$ to $\Tc = -1\deg$. In the third, we adopt a cylindrical tank instead of the 
original cuboidal counterpart. As shown in \cref{Fig Isothermal tank,Fig warmer disc,Fig circular tank}, all three configurations exhibit symmetry breaking at nearly the same time. The corresponding critical plume Rayleigh numbers $\Rac$ are $1.2 \times 10^6$, $1.3 \times 10^6$, and $1.9 \times 10^6$, respectively---all close to the threshold identified for the baseline case.
These findings  demonstrate that the identified threshold is a robust indicator for symmetry breaking and  spontaneous disk motion. 
The relatively large departure of $\Rac$ from the baseline observed in the third configuration (\cref{Fig circular tank}) may arise from the numerical representation of the circular tank's curved boundary. Specifically, the cylindrical wall is implemented via a single-node curved boundary condition~\citep{xiang2023lattice}, which involves extensive interpolation and thus introduces additional numerical errors.

\subsection{Amplitude of Wave Front}
\label{Front Amplitude}

\begin{figure}[tbhp!]
\begin{center}
\includegraphics[scale=1.0]{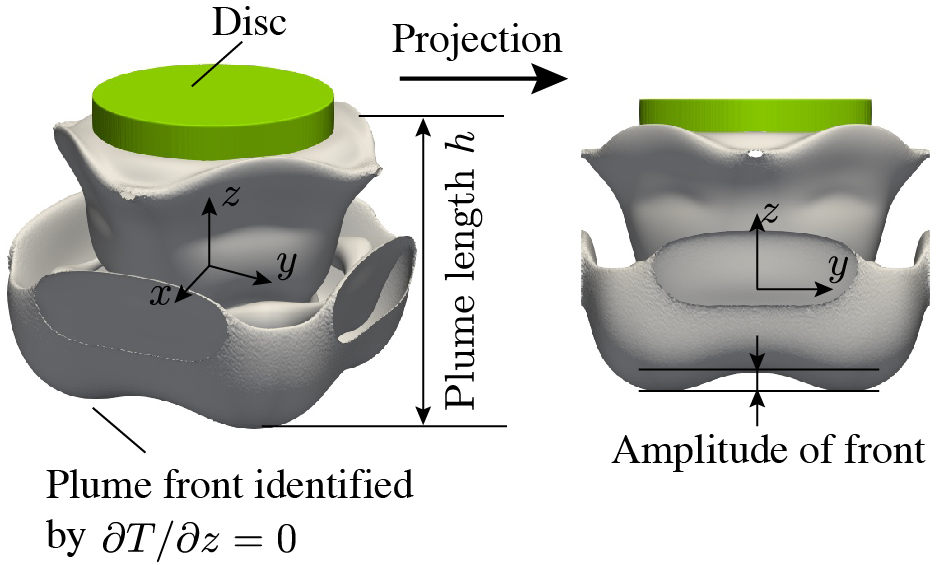}
\end{center}
\bfcaption{Amplitude of plume front}{
Illustration from the baseline simulation (\cref{Fig moving discs}) at dimensionless time $\tt=8.7$. 
}
\label{Fig frontamplitude}
\end{figure}

In the main text, we use the amplitude of the wavy front to characterize the linear and nonlinear stages of the RT instability \cref{LSA growth}.
To  measure this amplitude, we first identify the plume front  where $\partial T/\partial z=0$.
This front is then projected onto the vertical midplane ($yz$ plane), without loss of generality (see \cref{Fig frontamplitude}).

\end{document}